\documentstyle{article}
\newcommand{\lsim}{\:\raise -4pt\hbox{$\stackrel{\textstyle <} {\sim}$}\:}

\begin{document}
\begin{titlepage}
\begin{flushright}
Report-no: BTHEP-TH-96-42\\
Dec. 1996
\end{flushright}

\begin{center}
\LARGE
{\bf A New Approach of $J/\psi$ suppression in pA and AA Collisions}\\
\vspace{0.3in}
\large
Tai An$^{b}$ \footnote{taian@hptc1.ihep.ac.cn }, Chao Wei Qin$^{a,b} $
\footnote{chaowq@ccastb.ccast.ac.cn} and Yao Xiao Xia$^{b} $
\footnote{yxx@ccastb.ccast.ac.cn}

\begin{center}
\vspace{0.3in}\begin{tabbing}
ttttt \= tt \= \kill
\>a) CCAST (World Lab.), P. O. Box 8730 Beijing, P.R. China. \\
\>b) Institute of High Energy Physics, Academia Sinica, \\
\> \> P. O. Box 918(4-1), Beijing, 100039, P.R. China.\\      
\end{tabbing}
\end{center}

\vspace{0.3in}


\end{center}

\vspace{0.4in}
\normalsize
\begin{abstract}
When the cross section of $J/\psi$ production is considered varying with the energy of the
nucleon-nucleon interaction the production of $J/\psi$ in pA and AA collisions
has been studied using FRITIOF Model. The calculation shows that the cross
section of $J/\psi$ production ``per nucleon-nucleon collision" decreases  with  
increasing mass number
and centrality as a consequence of continuous energy loss of  the projectile 
nucleons to the target nucleons in their successive binary nucleon-nucleon collisions. 
We have compared our
model predictions with the experimental data of $J/\psi$ production.

\end{abstract}
\end{titlepage}
\normalsize
Suppression of $J/\psi$ production in high energy heavy ion collisions was
proposed as an effective signature of QGP formation ten years ago \cite{satz1}.
The ensuing experimental data  confirmed a significant suppression of 
$J/\psi$ production in both pA and AA collisions \cite{NA381}. However, alternative
explanations of the $J/\psi$ suppression exist based on the absorption of $J/\psi$
in nuclear matter \cite{capella}. The overall set of extensive data collected
and analysed by NA38 seems to support the absorption mechanism in p+B$_{target}$,
O+B$_{target}$ and S+U collisions. Nevertheless a rather large absorption cross
section ($\sim$ 6.2 mb) has to be used in order to fit the experimental data, about
three times larger than the total $J/\psi$-N cross section from the EMC
Collaboration \cite{JJ}. Such a difference was already noted long time ago in 
\cite{peng}. Recent 
calculations based on the colour octet model show that this phenomenological cross
section could be understood by the absorption of pre-resonance states ($ccg$) 
in nuclear matter \cite{Satz3}. Other sources of  $J/\psi$ suppression 
in nuclear collisions, such as
the interaction of $J/\psi$ particles with the produced mesons (called $comover$) 
\cite{Gavin}, 
gluon shadowing in nuclei, intrinsic charm component, energy degradation of produced
$c\bar{c}$ pair, etc. \cite{Vogt}, have been  introduced besides the absorption due to 
the  $J/\psi$-N interaction to explain the data.  

We in this paper propose a simple model to investigate  $J/\psi$ suppression 
in pA and AA collisions focusing on the decrease of the cross section of $J/\psi$
production with increasing mass number and centrality due to the continuous 
energy loss of  the projectile 
nucleons to the target nucleons in their successive binary nucleon-nucleon collisions,
 not on its later
absorption in nuclear matter. From the calculations of this model, we conclude
that the absorption of $J/\psi$ particles in nuclear matter is only part 
of the sources of $J/\psi$ suppression seen in pA and AA collisions so far.

Since the probability for a nucleon-nucleon collision leading to $J/\psi$ production
is very small it is generally accepted that multiple  $J/\psi$ production 
processes in pA and AA collisions can be neglected. The studies of 
$J/\psi$ suppression based on the absorption mechanism actually assume that
the probability of $J/\psi$ production ``per nucleon-nucleon collision" is the
same independent of the masses of the colliding nuclei at a given energy. 
However, each binary nucleon-nucleon collision experienced by a projectile 
nucleon on its way out of the target in pA and AA collisions may  not be the
same if the projectile nucleon loses a fraction of its energy in each
binary collision to the target nucleon such that the cms energy of each 
binary collision of this incoming nucleon with a target nucleon is different.
For the cross sections of the hard QCD parton-parton scatterings, which is
responsible for $J/\psi$ production, increase with  increasing energy it is
conceivable that the probability of $J/\psi$ production would also depend
on the energy of a nucleon-nucleon collision in a similar way. It is confirmed
both theoretically and experimentally that $J/\psi$ photoproduction
cross section exhibits a strong threshold behaviour at the low energy region and
then increases with energy at the higher energy region \cite{add1}.

In a participant-spectator model of nucleus-nucleus collisions, like FRITIOF,
each projectile nucleon may collide several times with the target nucleons.
If momentum transfers are assumed to take place in each binary nucleon-nucleon
collision, then the cms energy of these binary collisions will decrease with
time, thereby the probability to produce $J/\psi$  in each binary collision
going down.

We in this paper have calculated how the cross section of $J/\psi$ production
varies with increasing mass number of colliding nuclei and centrality using
FRITIOF model. The calculations show that $J/\psi$ production is suppressed in
pA and AA collisions in comparison with the nucleon-nucleon collision, and the
larger the mass number of colliding nuclei and the centrality,  the greater
the suppression,  as a result that probability of finding a  QCD hard 
process in a binary nucleon-nucleon decreases with increasing mass number 
of colliding nuclei and centrality. Taking into account the $J/\psi$
absorption by $J/\psi$-N interaction our results are in good agreement with 
experimental data with the exception of the latest data of Pb$+$Pb collisions
at 158 AGeV/c, which show a further $J/\psi$ suppression \cite{bin}. 
The absorption cross section $\sigma_{abs}$
needed to explain the data  from p+B$_{target}$, O+B$_{target}$ and 
S+U collisions is about 1.4 mb in this paper, which is consistent with the 
experiments (EMC
Collaboration \cite{JJ} gives a total $J/\psi$-N cross section 2.2 $\pm$ 0.7 mb
and the $J/\psi$-N quasi-elastic cross section is given in \cite{quasi} as
0.79 $\pm$ 0.012 mb). Furthermore, our results also imply that some new
mechanism of $J/\psi$ suppression seems to be needed particularly 
for understanding 
Pb$+$Pb data. The authors in \cite{JP} \cite{wong} have attributed
the further $J/\psi$ suppression in Pb+Pb data to the formation of a QGP state.


As we have mentioned before the cross section of $J/\psi$ production
will be different in each binary nucleon-nucleon collision in pA and AA
collisions if the 
energy loss of the projectile nucleons to the target nucleons in the successive 
binary 
collisions is taken into account. Assume that $<\sigma^{NN}_{J/\psi}>$ is the 
mean cross section for the production of a $J/\psi$ particle in a binary
nucleon-nucleon collision (here the average is done over all the binary
collisions at an impact parameter $b$), 
then the total probability for producing a $J/\psi$ particle
in the collisions of A + B at an impact parameter $b$ is the sum \cite{wong2}
\begin{equation}
P^{AB}_{J/\psi} = \sum^{AB}_{n=1}\left(\begin{array}{c}
AB\\
n\\
\end{array} \right)[T(b)<\sigma^{NN}_{J/\psi}>]^n[1-T(b)<\sigma^{NN}_{J/\psi}>]
^{AB-n},
\label{ff1}
\end{equation}
where T(b) is the thickness function. Because $T(b)<\sigma^{NN}_{J/\psi}>$ is a very
small quantity ($<\sigma^{NN}_{J/\psi}> \sim 10^{-4}$ fm$^2$ \cite{sigma}, $T(b) \sim
10^{-2} $fm$^{-2}$ for central S+U collisions, for instance), the summation given by 
Eq.(\ref{ff1}) is dominated by the first term with n=1. The terms with
n$ >$1 represent multiple $J/\psi$ production processes and shadowing 
corrections, which are very small and can be neglected. Then the 
probability for $J/\psi$ production in A+B collisions can be approximated
to be
\begin{equation}
P^{AB}_{J/\psi} = AB[T(b)<\sigma^{NN}_{J/\psi}>].
\label{ff2}
\end{equation}
Therefore, the cross section of $J/\psi$ production corresponding to
a centrality bin, $\Delta b$, is given by the following formula
\begin{equation}
\sigma^{AB}_{J/\psi,\Delta b} = AB<\sigma^{NN}_{J/\psi}>_{\Delta b}
\int_{\Delta b}T(b)db,
\label{ff3}
\end{equation}
where $<\sigma^{NN}_{J/\psi}>_{\Delta b}$ is the mean cross section of $J/\psi$ 
production ``per nucleon-nucleon collision'' within the centrality bin $\Delta b$. 
We know that QCD hard scatterings (the gluon fusion and quark-antiquark
annihilation) between partons are the main source of $J/\psi$ production.
 Let $P_{h}$ be the probability to have a hard scattering
in a nucleon-nucleon collision and  $P_{h}^{J/\psi}$ the probability to produce
a $J/\psi$ from the hard scattering, then we can write out $<\sigma^{NN}_{J/\psi}>$
to be
\begin{equation}
<\sigma^{NN}_{J/\psi}>=<\sigma_{T}P_{h}^{J/\psi} P_{h}>=
\sigma_{T}P_{h}^{J/\psi} <P_{h}>,
\label{ff4}
\end{equation}
where $\sigma_{T}$ is the total cross section of a nucleon-nucleon collision.
We have assumed that $\sigma_{T}P_{h}^{J/\psi}$ is approximately a constant
in the energy span that we are concerning, so the product is the same for all
the binary collisions. Combining Eq.(\ref{ff4}) with 
Eq.(\ref{ff3}) and replacing $<P_{h}>$ by the ratio $\frac{n^{h}_{bin}(b)}
{n_{bin}(b)}$
($n_{bin}(b)$ is the number of binary collisions in an A+B collision and 
$n^{h}_{bin}(b)$ the number of binary collisions with a hard scattering. Both
of them are the function of the impact parameter $b$) we finally
obtain
\begin{equation}
\sigma^{AB}_{J/\psi,\Delta b} = AB\sigma_{T}P_{h}^{J/\psi} 
<\frac{n^{h}_{bin}(b)}{n_{bin}(b)}>_{\Delta b}\int_{\Delta b}T(b)db
\label{ff5}
\end{equation}
and for the minimum bias events we have
 \begin{equation}
\frac{\sigma^{AB}_{J/\psi}}{AB} = \sigma_{T}P_{h}^{J/\psi} \frac{n^{h}_{bin}}
{n_{bin}}.
\label{ff6}
\end{equation}
We see from Eq.(\ref{ff6}) that the dependence of the quantity 
$\frac{\sigma^{AB}_{J/\psi}}{AB}$ on the masses of colliding nuclei or centrality
is solely determined by how the mean probability of having a hard 
scattering in a binary
nucleon-nucleon collision varies with the masses of colliding nuclei or centrality.
Before calculating $\frac{n^{h}_{bin}(b)}{n_{bin}(b)}$ 
in pA and AA collisions we will give a brief introduction of FRITIOF dynamics focusing on how
a hard parton-parton scattering is distinguished from a soft one.

FRITIOF is a string model based on the concepts of the Lund String Model
\cite{BAphyRep}, which started from the modeling of inelastic 
hadron-hadron collisions and it has
been successful in describing many experimental data from the low energies
at the ISR-regime all the way  to the top SPS energies \cite{B.N} \cite{H.P1}. 
This has been achieved by the introduction of a
particular longitudinal momentum transfer scenario, gluon
bremsstrahlung radiation (The Dipole Cascade Model, DCM \cite{DCM}, and the Soft
Radiation Model, SRM \cite{SRM}, --- this is
implemented by the use of ARIADNE \cite{aria}) as well as hard parton
scattering (Rutherford Parton Scattering, RPS --- this is implemented
by the PYTHIA routines \cite{PYT}). In FRITIOF, during the collision
two hadrons are excited due to longitudinal momentum transfers and/or a
RPS. It is further assumed that there is no
net color exchange between the hadrons. The highly excited states will emit bremsstrahlung gluons according
to the SRM. They are afterwards treated as excitations or the Lund Strings and 
the string states are
allowed to decay into final state hadrons according to the Lund
prescription as implemented by JETSET \cite{jet} .

In the FRITIOF model a hadron is assumed to behave like a massless
relativistic string (MRS) corresponding to a confined color force field
of a vortex line character embedded in a type II color superconducting
vacuum. A hadron-hadron collision is pictured as the multi-scatterings 
of the partons  inside the two colliding hadrons.
This includes both the hard and the soft components depending on the four-momentum
transfers $Q^2$, or equivalently the transverse momentum transfers involved. 
The soft part is described by a simple phenomenological model.
The hard scatterings
can however be calculated from perturbative QCD, and correspond to the 
Rutherford parton-parton scattering (RPS).  The divergence problem in RPS is handled
by introducing the Sudakov factor.

There will be color separation in the model, i.e. there will
for each hadron be a color $\bar{3}$ (a ``diquark'') continuing forward
along the beam direction and a valence quark, a color $3$, moving in
the opposite direction due to the longitudinal momentum transfer.
This will lead to bremsstrahlung of a dipole character.

A procedure  therefore is adopted in FRITIOF that compares
the ``hardness'' of the Rutherford partons to that of the bremsstrahlung gluons.  
The RPS is accepted only if
it is harder than the associated radiation.  If the RPS is ``drowned'', 
which is to say that it is softer than the radiation, then the RPS is not
acceptable and the collision proceeds as a purely soft collision. 
With this prescription the RPS spectrum is suppressed smoothly at small
to medium transverse momentum region. 

For the hadron-nucleus and nucleus-nucleus collisions, the process has
in the FRITIOF model been treated as a set of incoherent collisions on the
nucleons. Thus a nucleon from the projectile interacts independently with the
encountered target nucleons as it passes through the nucleus. The probability 
distribution for the number of inelastic collisions $\nu$ is taken from
geometric calculations. Each of the sub-collisions is treated in the
same way as an ordinary hadron-hadron collision, although the momentum
transfers will again be additive and every encounter will make the
projectile more excited. If it interacts with $\nu$ nucleons in the target, $\nu+1$ 
excited string states will be formed as a result. These string states 
will then independently emit associated bremsstrahlung radiation
and then fragment into hadrons in the same way as individual strings.
This picture is supported by the fact that the global features of heavy ion 
collisions are satisfactorily explained by the collision geometry 
together with the independent hadron-hadron collisions.

Using FRITIOF it is straightforward to calculate the number of binary nucleon-
nucleon collisions, $n_{bin}(b)$, and the number of the binary collisions with a hard 
scattering, $n_{bin}^{h}(b)$, in pA and AA collisions at a given 
impact parameter $b$, so that the mean probability to have a hard scattering 
in a binary nucleon-nucleon collision,  $<P_{h}>$ = $\frac{n_{bin}^{h}(b)}{n_{bin}(b)}$, 
can be obtained. Since we are mainly interested in  $J/\psi$ production in pA and
AA collisions relative to that in the pp collision we do not need to know how 
a $J/\psi$ is actually formed from the hard scatterings in order to investigate 
the dependence of  $J/\psi$ production cross sections on mass number and centrality.

We have calculated the quantity $\frac{n^{h}_{bin}}{n_{bin}}$ for 
various pA and AA
collisions (and $<\frac{n^{h}_{bin}(b)}{n_{bin}(b)}>_{\Delta b}$ for different 
centrality bins in S+U and Pb+Pb collisions)
 at $P_{lab}$=200 GeV/c using FRITIOF. After determining 
$\sigma_{T}P_{h}^{J/\psi}$ by the data of the pp collision we plot our results 
of $B_{\mu\mu}\frac
{\sigma^{AB}_{J/\psi}}{A_{proj.}\cdot B_{targ.}}$ as a function of $A_{proj.}\cdot 
B_{targ.}$ in Figure 1
for the minimum bias events.  For the cross section of  $J/\psi$ production in
different centrality bins we plot the results of $B_{\mu\mu}\frac
{\sigma^{AB}_{J/\psi}}{\sigma_{DY}}$ as a function of $N_{p}\cdot N_{t}$ 
in Figure 2, where $N_{p}$ is the number of participants from the  projectile and $N_{t}$ 
the number 
of participants from the target, since the Drell-Yan cross section in a given
centrality bin is found
in experiments proportional to an effective $A_{proj.}\cdot B_{targ.}$
(= $A_{proj.}\cdot B_{targ.}\int_{\Delta b}T(b)db$). In the same way, a constant
has to be determined by the corresponding data of the pp collision. The impact parameter bins are 
taken to be the same as 
those extracted by NA38 and NA50 \cite{bin}. We decided not to use the absorption 
length $L$ to be the  longitudinal axis as used by NA50 because $L$  calculated from 
the geometry model is not sensitive
to the change of impact parameter for very central Pb+Pb collisions.
The results of our calculations show that the 
decrease of quantity, $B_{\mu\mu}\frac{\sigma^{AB}_{J/\psi}}{A_{proj.}\cdot B_{targ.}}$ or  $B_{\mu\mu}\frac
{\sigma^{AB}_{J/\psi}}{\sigma_{DY}}$ (the cross section of $J/\psi$ production ``per
nucleon-nucleon collision"), is due to the fact that the probability of the QCD hard
scattering per binary nucleon-nucleon collision decreases with the increasing 
mass number and centrality.

When the absorption of $J/\psi$ by $J/\psi$-N interaction is also taken into account, i.e. the
previous results are multipied by $\exp(-\rho L \sigma_{abs})$ with $\rho$=0.14 n/fm$^3$, $
 \sigma_{abs}$=1.4 mb  and $L$ taken to be the same as those in \cite{bin}, our model
 reproduces the data of $J/\psi$ suppression with the exception of the latest
 data from Pb+Pb collisions at 158 GeV/c, which clearly show a further suppression.
 
 One possibility which will bring about  a further suppression of $J/\psi$ production
is that $P^{J/\psi}_{h}$, the probability to produce a  $J/\psi$ from a hard process, 
drops down suddenly under  certain conditions. This is equivalent to say that the
$c\bar{c}$ produced from the hard processes can not form a bound state. However at the
moment our simple model can not estimate when this would happen.

However, there are still other possible mechanisms of $J/\psi$ suppression, which
are not included in our simple model. The $x_{F}-$ dependence of 
$J/\psi$ suppression is not investigated yet. Therefore, it is hard to make any 
conclusion now whether this further suppression in Pb+Pb collision is due to
QGP formation.
 
 It is known that there is no unique criterion to distinguish a hard process from
 a soft one in a nucleon-nucleon collision. Usually a $q_{Tmin}$ is introduced to 
 be the minimum  transverse momentum of the produced partons from a hard
 process. A dynamic criterion is applied in FRITIOF to chose a hard process by
 comparing the hardness of a RPS parton with the hardness of the bremsstrahlung gluons
 as mentioned before. However, the cross section of   $J/\psi$ production should
 not depend on which criterion is actually used in the calculation. We have thus
 calculated all the results of this paper using the conventional $q_{Tmin}$ criterion
 in Pythia ($q_{Tmin}$= 1GeV/c), just to check if our conclusion relies on the specific
 criterion in FRITIOF. The calculations show that the results in these two cases
 are in agreement with each other.
 
 We have also checked if the quantity $\sigma_{T}P^{ J/\psi}_{h}$ is energy-independent  
 as we have assumed. A parametrization form of the $J/\psi$ cross section is given as
 \cite{param}
\begin{equation}
\sigma^{J/\psi} = \sigma^0(1-\frac{M_{J/\psi}}{\sqrt{s}})^{12},
\label{ff7}
\end{equation}
where $\sqrt{s}$ stands for the cms energy per nucleon. Therefore, if 
$\sigma_{T}P^{ J/\psi}_{h}$ in Eq.(\ref{ff6}) is energy-independent then we should have
a ratio
\begin{equation}
\frac{(1-\frac{M_{J/\psi}}{\sqrt{s}})^{12}}{\frac{n^{h}_{bin}}{n_{bin}}}= C =\mbox{ constant.}
\label{ff8}
\end{equation}
We have calculated $\frac{n^{h}_{bin}}{n_{bin}}$ 
 
and the ratio at various energies from $P_{lab}$=60 GeV/c to $P_{lab}$=
450 GeV/c for pp collisions ($n_{bin} $= 1 for a pp collision) 
and the results are listed in Tab.1, which show that this
ratio in this energy region is not sensitive to the change of energy in 
comparison with 
${\frac{n^{h}_{bin}}{n_{bin}}}$. But a threshold behaviour may exist at lower 
energies, which can be seen from the value at  $P_{lab}$=60 GeV/c.

 \section{Acknowledgments}
We would like to thank D. Jouan and Zhang Xiaofai for helpful discussions. 
This work is partly supported by the National Natural Science Founction of China.


 \newpage
{\bf Figure Captions}\\

Figure1: $J/\psi$ cross sections divided by $A_{proj.}\cdot 
B_{targ.}$ as a function of $A_{proj.}\cdot B_{targ.}$. Our results
are compared with the experimental data  from NA51, NA38 and NA50 \cite{bin}. \\

Figure2: $J/\psi$ cross sections divided by Drell-Yan cross sections
  as a function of $N_{p} \cdot N_{t}$. Our results
are compared with the experimental data  from NA38 and NA50 \cite{bin}.

 \newpage
 \vskip 0.3cm
\large
\begin{center}
 \begin{tabular}{||c|c|c||}  
 \multicolumn{3}{c}{Table 1. The energy dependence of the 
 probability to} \\
 \multicolumn{3}{c}{have a hard scattering per binary collision} \\
\multicolumn{3}{c}{and the assumed constant C in Eq.(\ref{ff8})}\\  
\hline  \hline
 $P_{lab}\ $ GeV/c & $\frac{n^{h}_{bin}}{n_{bin}}$&$C$ \\
 \hline
  $60$&$1.17\times 10^{-2}$&$1.84$  \\
 \hline
  $100$&$2.82\times 10^{-2}$&$2.03$  \\
 \hline
  $150$&$4.76\times 10^{-2}$&$2.14$  \\
 \hline
  $200$&$6.40\times 10^{-2}$&$2.22$  \\
  \hline
  $300$&$9.17\times 10^{-2}$&$2.27$  \\
 \hline
  $450$&$12.2\times 10^{-2}$&$2.30$  \\ 
\hline\hline
 \end{tabular}
 \end{center}  
                  

\newpage
\begin{thebibliography}{35}
\bibitem{satz1}T. Matsui and H. Satz, Phys. Lett. {\bf B178}, 416 (1986).
\bibitem{NA381}NA38 Collaboration, C. Baglin et al., Phys. 
Lett. {\bf B220}, 471 (1989); {\bf B251}, 465 (1990).
\bibitem{capella}A. Capella et al., Phys. Lett. {\bf B206}, 47 (1988); C. Gerschel and J. 
Huefner, Phys. Lett. {\bf B207}, 253 (1988).
\bibitem{JJ} J. J. Aubert et al., Nucl. Phys. {\bf B213}, 1 (1983).
\bibitem{peng} J.-C. Peng et al., in Proceedings of Workshop on Nuclear Physics
on the Light Cone, July 1988, World Scientific Publisher, P.65.
\bibitem{Satz3} D. Kharzeev and H. Satz. Phys. Lett. {\bf B366}, 316 (1996).
\bibitem{Gavin} S. Gavin and R. Vogt. Nucl. Phys. {\bf B345}, 104 (1990).
\bibitem{Vogt} R. Vogt, S. J. Brodsky and P. Hoyer, Nucl. Phys. {\bf B360},
 67 (1991); R. Vogt, Nucl. Phys. {\bf A544}, 615c (1992); 
 W. Q. Chao and B. Liu, Z. Phys. {\bf C72}, 291 (1996).
\bibitem{add1} D. Kharzeev, H. Satz, A. Syamtomov and G. Zinoviev, 
CERN-TH/96-72.
\bibitem{bin} NA50 Collaboration to appear in 
Proceedings of Quark Matter'96.
\bibitem{quasi} R. L. Anderson, SLAC-Pub 1741 (1976).
\bibitem{JP} J.-P. Blaizot and J. Y. Ollitrault, Phys. ReV.  Lett. {\bf 77}, 
1703 (1996).
\bibitem{wong} C.-Y. Wong, ORNL-CTP 96/07 (1996).
\bibitem{wong2} C.-Y. Wong, `` Introduction to High Energy Heavy-Ion Collisions", 
World Scientific publishing, Singapore 1994, P.360.  
\bibitem{sigma} N. S. Craigie, Phys. Rep. {\bf 47}, 1 (1978).  
\bibitem{BAphyRep} B. Andersson, G. Gustafson, G. Ingelman and T. Sj\"{o}strand, 
Phys. Rep. {\bf 97}, 31 (1983).
\bibitem{B.N} B. Andersson, G. Gustafson and B. Nilsson-Almqvist, 
Nucl. Phys. {\bf B281}, 289 (1987).
\bibitem{H.P1}  B. Andersson, G. Gustafson and H. Pi, Z. Phys. {\bf C57}, 485 (1993).
\bibitem{DCM} G. Gustafson, Phys. Lett. {\bf B175}, 453 (1986); G. Gustafson 
and U. Pettersson, Nucl. Phys. {\bf B306}, 746 (1988).
\bibitem{SRM} B. Andersson et al. Z. Phys. {\bf C43}, 625 (1989).
\bibitem{aria} L.\ L\"onnblad, 
  ``{\tt ARIADNE version 4}, A Program for Simulation of QCD Cascades 
implementing the Color Dipole Model'', DESY 92-046.
\bibitem{PYT} H. -U. Bengtsson and T. Sj\"ostrand, Comput. Phys. Commun. 
{\bf 46}, 43 (1987).
\bibitem{jet} T. Sj\"ostrand, 
   ``A Manual to The Lund Monte Carlo for Jet Fragmentation and
     $e^+e^-$ Physics: JETSET version 7.3'', available upon request
     to the author.
\bibitem{param} G. A. Schuler, CERN-TH 7170/94.
\end{thebibliography}
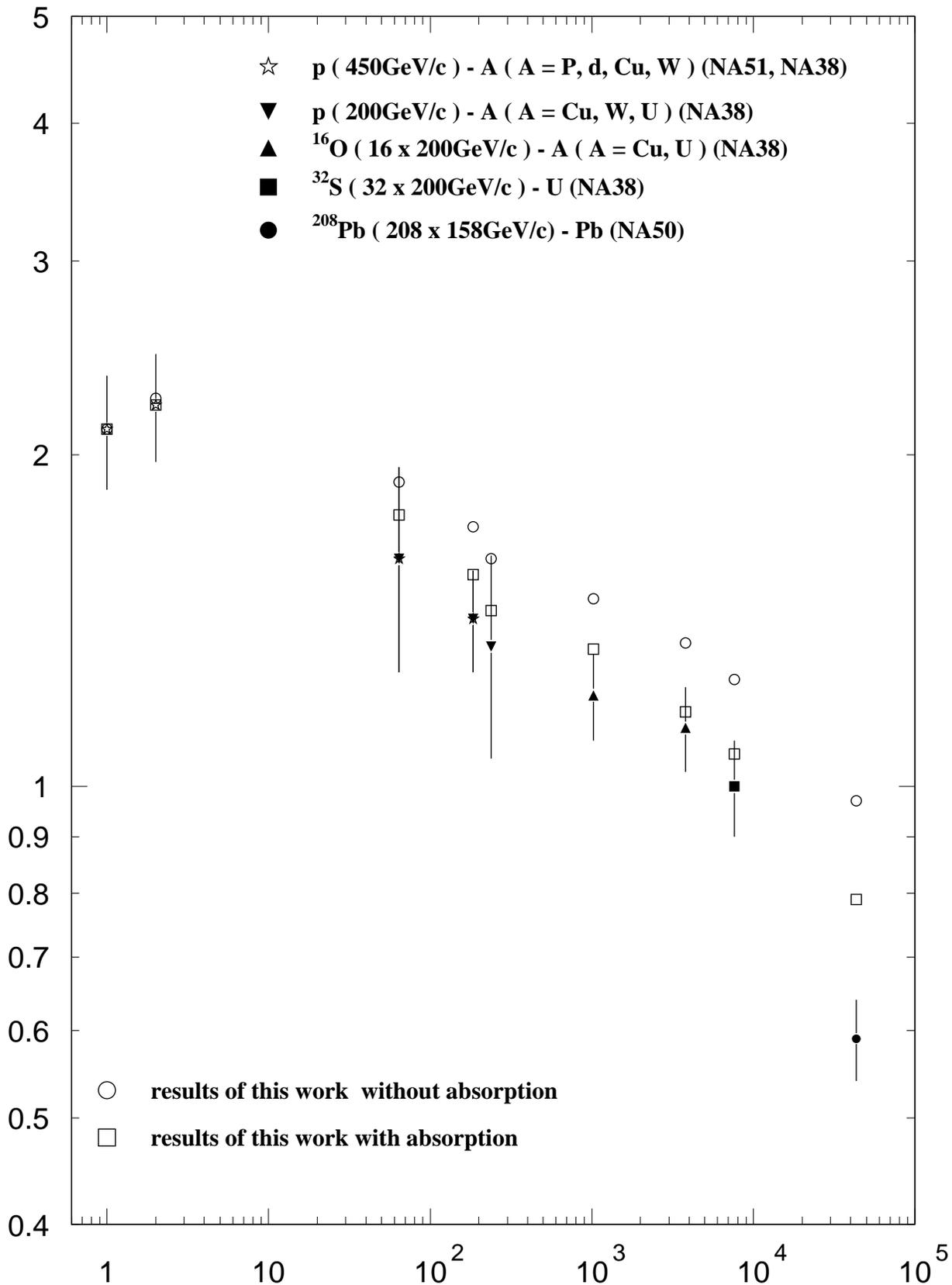
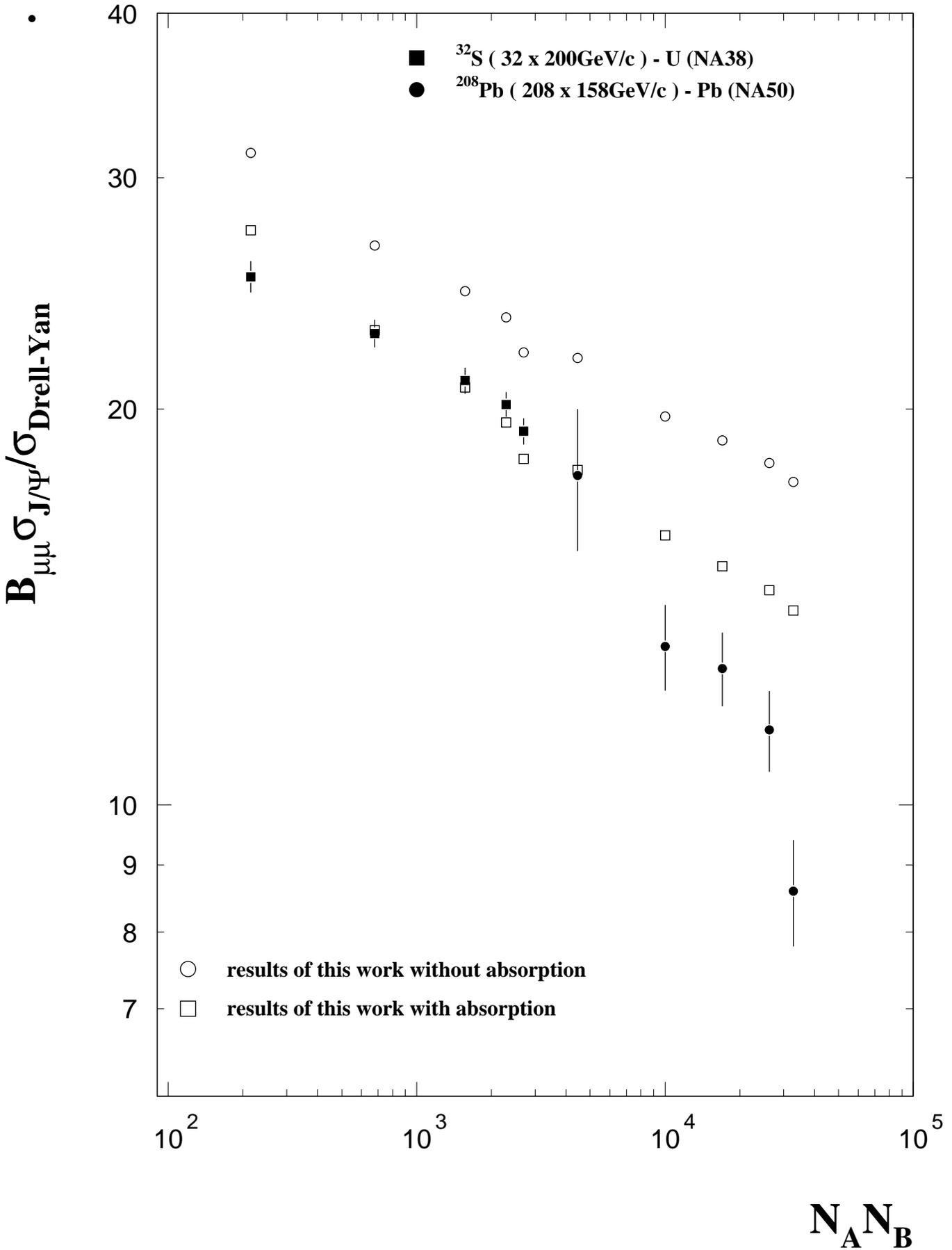
\end{document}